\documentclass[twocolumn,showpacs,preprintnumbers,amsmath,amssymb]{revtex4}
\usepackage{graphicx}
\usepackage{dcolumn} 
\usepackage{bm}      

\begin{document}
\title{TMCI threshold with space charge and different wake fields}
\author{V. Balbekov}
\email{balbekov@fnal.gov}
\affiliation {Fermi National Accelerator Laboratory\\
 P.O. Box 500, Batavia, Illinois 60510, email balbekov@fnal.gov} 
\date{\today}

\begin{abstract}

Transverse mode coupling instability of a bunch with space charge and wake 
field is considered within the frameworks of the boxcar model.
Eigenfunctions of the bunch without wake are used as a basis for the solution 
of the equations with the wake field included.
A dispersion equation for constant wake is presented in the form of 
an infinite continued fraction and also as the recursive relation with an
arbitrary number of the basis functions.
Realistic wake fields are considered as well including resistive wall, square, 
and oscillating wakes.
It is shown that the TMCI threshold of the negative wake grows up in absolute 
value when the SC tune shift increases.
Threshold of positive wake goes down at the increasing SC tune shift.
The explanation is developed by an analysis of the bunch spectrum.
	
\end{abstract}

\maketitle
%

\section{INTRODUCTION}

%
Transverse mode coupling instability (TMCI) has been observed first in PETRA 
and explained by Kohaupt on the base of the two-particle model \cite{KO}. 
A lot of papers on this subject have been published later, 
including handbooks and surveys (see e.g. \cite{NG}).
It is established that the instability occurs as a result of a coalescence
of the neighboring head-tail tunes caused by the bunch wake field.  

TMCI with space charge has been considered first by Blaskiewicz \cite{BL1}.
The main point of this paper is that the SC pushes up the TMCI threshold 
that is improves the beam stability.
However, a non-monotonic dependence of the TMCI threshold and rate on the SC
tune shift has been sometimes demonstrated in the paper.
It was following from  several examples that the stability and instability areas 
can change each other when the tune shift increases.
The results have been confirmed later by the same author with help of 
numerical simulation of the instability at modest magnitude of the 
SC tune shift \cite{BL2}

So-termed three-mode model has been developed in Ref.~\cite{BA2} for analytical 
description of the TMCI with space charge, chromaticity, and arbitrary wake.
This simple model confirms that the TMCI threshold of negative wakes goes up 
in modulus when the SC tune shift increases.
However, only the case of modest SC has been investigated in \cite{BA2} 
though the proposed equations allow to suggest that a sudden kink of the threshold 
curve is possible at the higher shift.
Therefore, field of application of the three-mode model is still an open question.

The case of very high space charge has been considered in Ref.~\cite{BUR, BA1}
It was confirmed in both papers that the space charge heightens the TMCI threshold
until the ratio of the SC tunes shift to synchrotron tune is less of
several tens or a hundred.
However, the authors have expressed the different opinions about further 
behavior of the threshold.
As it follows from \cite{BA1}, the threshold growth should continue at 
higher SC as well.   
On the contrary, it was suggested in Ref.~\cite{BUR} that the threshold growth 
can cease and turn back over the mention boundary.

The last statement has been supported in my recent preprint \cite{BA3}.
I have used the known eigenfunctions of the boxcar bunch \cite{SAH}  
to get a convenient basis for investigation of the TMCI problem in depth. 
However, disclosure of some errors at numerical solutions of obtained 
equations forces me to revise the conclusions.
The equations are recomputed in presented paper at any value of the SC tune 
shift and different wakes including the resistive wall, square, the 
oscillating ones. 
The pushing up of the TMCI threshold by the SC is observed in all the cases.

%

\section{BASIC EQUATIONS and ASSUMPTIONS}

%

The terms, basic symbols and equations of Ref.~\cite{BA2}
are used in this paper.
In particular, linear synchrotron oscillations are considered here
being characterized by amplitude $\,A\,$ and phase $\,\phi$, 
or by corresponding Cartesian coordinates: 
%
\begin{equation}
 \theta= A\cos\phi, \qquad u=A\sin\phi.
\end{equation}
%
Thus $\,\theta\,$ is the azimuthal deviation of a particle from the bunch center
in the rest frame, and variable $u$ is proportional to the momentum 
deviation about the bunch central momentum 
(the proportionality coefficient plays no part in the paper).
Transverse coherent displacement of the particles in some point of the 
longitudinal phase space will be presented as real part of the function 
%
\begin{equation}
 X(A,\phi,t)=Y(A,\phi)\exp\big[-i(Q_0+\zeta)\,\theta-i\,(Q_0+\nu)\,\Omega_0t\,\big]
\end{equation}                         
%
where $\,\Omega_0\,$ is the revolution frequency, $\,Q_0$ is the central 
betatron tune,  and $\,\nu\,$ is the tune addition produced by space charge and 
wake field. 
Generally, $\,\zeta\,$~is the normalized chromaticity, however, only the case 
$\zeta=0$ will be investigated in this paper. 
Then the function $Y$ satisfies the equation \cite{BA1},\cite{BA2}:
%
\begin{eqnarray}
 \nu Y+i\,Q_s\frac{\partial Y}{\partial\phi}
+\Delta Q(\theta)\big[Y(\theta,u)-\bar Y(\theta)\big]
 \nonumber \\
=2\int_\theta^\infty q(\theta'-\theta)\bar Y(\theta')\rho(\theta')d\theta' 
\end{eqnarray}
%
where $\,F(\theta,u)\,$ and $\,\rho(\theta)\,$ are the normalized distribution 
function and corresponding linear density of the bunch, $\,Q_s\,$ is the 
synchrotron tune, $\,\Delta Q(\theta)\propto \rho(\theta)\,$ is the space charge 
tune shift, and $\,\bar Y(\theta)\,$ is the bunch displacement in usual space 
which can be found by the formula
%
\begin{eqnarray}
 \rho(\theta)\bar Y(\theta) = \int_{-\infty}^\infty F(\theta,u)Y(\theta,u)\,du.  
\end{eqnarray}
%
The function $\,q(\theta)\,$ is proportional to the usual transverse wake field
$\,W_1$
%
\begin{eqnarray}
 q = \frac{r_0RN_bW_1}{8\pi\beta\gamma Q_0}
\end{eqnarray}
%
with $\,r_0=e^2/mc^2\,$ as classic radius of the particle, $\,R\,$ as 
the accelerator radius, $\,N_b\,$ as the bunch population, $\,\beta\,$ and 
$\,\gamma\,$ as the normalized velocity and energy \cite{NG}. 

A solution of Eq.~(3) can be found by its expansion in terms of the 
eigenfunctions of corresponding homogeneous equation which is
%
\begin{eqnarray}
 \nu_jY_j+i\,Q_s\frac{\partial Y_j}{\partial\phi}+\Delta Q(\theta)
 \,\big[Y_j(\theta,u)-\bar Y_j(\theta)\big]=0.
\end{eqnarray}
%
It is easy to check that the functions form an orthogonal basis with the weighting 
function $\,F(\theta,u)\,$. 
Besides, we will impose the normalization condition:
%
\begin{eqnarray}
 \int\int F(\theta,u)Y^*_j(\theta,u)Y_k(\theta,u)\,d\theta du=\delta_{jk}
\end{eqnarray}
%
where the star means complex conjugation.
Then, looking for the solution of Eq.~(3) in the form
%
\begin{equation}
 Y = \sum_j C_jY_j,
\end{equation}
%
one can get the expression for the unknown coefficients $\,C_j$:               
%
\begin{equation}
 \sum_j\,(\nu-\nu_j)\,C_jY_j = 2\sum_j C_j\int_{\theta}^\infty 
 \bar Y_j(\theta')\,\rho(\theta')\,q(\theta'-\theta)\,d\theta'
\end{equation}
%
where $\,\bar Y_j\,$ and $\,Y_j\,$ are also connected by Eq.~(4). 
Multiplying Eq.~(9) by factor $\,F(\theta,u)\,Y^*_J(\theta,u)\,$,
integrating over $\theta$ and $u$, and using normalization condition (7)
one can get the series of equations for the coefficients $\,C_j$:   
%
\begin{eqnarray}
 (\nu-\nu_J)C_J = 2\sum_j C_j\int_{-\infty}^\infty \rho(\theta)\,
 \bar Y_J^*(\theta) \,d\theta  \nonumber \\ \times\,
 \int_\theta^\infty\rho(\theta')\,\bar Y_j(\theta')\,q(\theta'-\theta)\,d\theta'.
\end{eqnarray}
%
%

\section{BOXCAR MODEL}

%

The boxcar model is characterized by following expressions for the bunch 
distribution function and its linear density:
%
\begin{subequations}
\begin{eqnarray}
 F = \frac{1}{2\pi\sqrt{1-A^2}}=\frac{1}{2\pi\sqrt{1-\theta^2-u^2}},
\end{eqnarray}
\begin{eqnarray}
 \rho(\theta)=\frac{1}{2}\qquad{\rm at}\qquad|\theta|<1.
\end{eqnarray}
\end{subequations}
%
Because the eigenfunctions depend on two variables $(\theta$-$u)$ (or $A$-$\phi)$,
it is more convenient to represent $\,j\,$ as a pair of the indexes: 
%
\begin{equation}  
 j \equiv \{n,m\},\qquad Y_j\equiv Y_{n,m}.
\end{equation}
%
An analytical solutions of Eq.~(6) for the boxcar bunch have been found 
by Sacherer~\cite{SAH}.
The most important point is that the averaged eigenfunctions $\,\bar Y_{n,m}\,$ 
do not depend on second index being proportional to the Legendre polynomials: 
$\bar Y_{n,m}(\theta) \propto 2\bar Y_n(\theta)\propto P_n(\theta),\;n=0,\,1,\,2,\dots$.
At any $\,n,$, there are $\,n+1\,$ different eigenmodes $\,Y_{n,m}(\theta,u)\,$
satisfying the equation
%
\begin{eqnarray}
 (\nu_{n,m}+\Delta Q)\,Y_{n,m}+i\,Q_s\frac{\partial Y_{n,m}}{\partial\phi}
=\Delta Q\,S_{n,m}P_n(\theta) 
\end{eqnarray}
%
where $\,m=n,\,n-2,\,\dots,\,-n$.
The space charge tune shift $\,\Delta Q\,$ is constant in this model, 
and the coefficients $\,S_{n,m}\,$ are added to the right-hand part 
of the equation to meet the normalization condition given by Eq.~(7).
Details of these calculations are placed in the Appendix, and several important
examples are represented in Figs.~1 and 2.
%
 \begin{figure}[t!]
 \includegraphics[width=85mm]{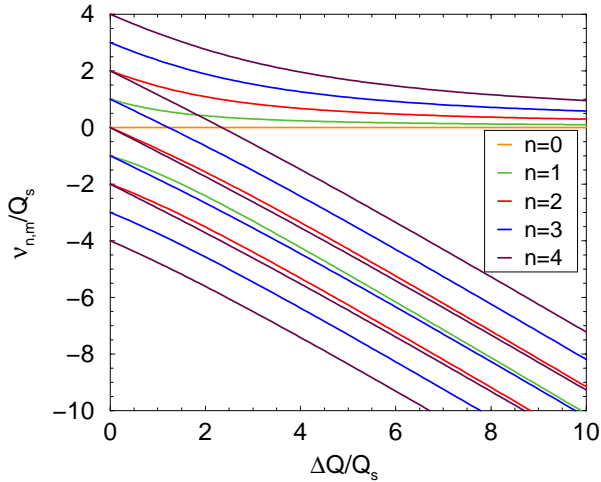}
 \caption{Eigentunes of the boxcar bunch without wake. 
 At any $\,n$, there are $\,n+1\,$ 
 eigentunes starting at $\,\Delta Q=0\,$ from the points $\,\nu_{n,m}=mQ_s$, 
 $\,m=n,\;n-2,\dots,-n$.}
 \end{figure}
%
 \begin{figure}[t!]
 \includegraphics[width=85mm]{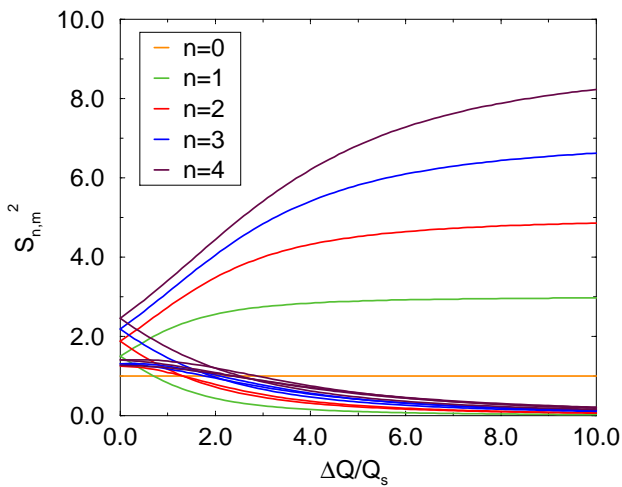}
 \caption{Normalizing coefficients of the boxcar bunch.
 The rising lines refer to the case $\,m=n\,$.}
 \end{figure}
%
Dependence of the eigentunes on the SC tune shift is shown in Fig.~1. 
It is seen that all of them take start at $\,\Delta Q=0\,$ from the points 
$\,\nu_{n,m}(0)=mQ_s$.
It is the commonly accepted convention to use the term ``multipole'' for the collective 
synchrotron oscillations of such frequency, 
that is the index $\,m\,$ should be treated here as the multipole number. 
Another index $\,n\,$ characterizes the eigenfunction power.
This feature is normally associated with a radial mode number,
the lower power corresponding to the lower number. 
Because $\,n\ge |m|$ in this case, the mode $\,\{m,m\}$ should be treated as the lowest radial 
mode of $\,m$-th multipole.  

At $\Delta Q\ne 0$, the multipoles mix together, and the eigentunes split on 2 groups. 
In the first of them, all tunes have positive value which tends to 0 at 
$\Delta Q/Q_s\rightarrow\infty$.
By the origin, all of them are the lowest radial modes $\,\{n,n\}$.
Corresponding normalizing coefficients $\,S_{m,m}^2\rightarrow 2n+1\,$ at 
$\Delta Q/Q_s\rightarrow\infty\,$ (Fig.~2).
In second group, the tunes are about 
$\nu_{n,m}\simeq mQ_s-\Delta Q\,$ being weakly dependent on the radial index $\,n$.  
The normalizing coefficients tend to 0 in this group.

Note that the transient conjugations of some eigentunes 
(the line crossing in Fig.~1) is not an evidence of instability in the case, 
because corresponding eigenfunctions are orthogonal and uncoupled.

With $\,\bar Y_{n,m}=S_{n,m}P_n(\theta),$ series (10) for the boxcar bunch 
obtains the form 
%
\begin{equation}
 (\nu-\nu_{N,M})C_{N,M}=q_0 S_{N,M}^*\sum_{n=0}^\infty R_{N,n}\sum_m S_{n,m}C_{n,m}
\end{equation}
%
with the matrix
%
\begin{equation}
 R_{N,n}=\frac{1}{2}
 \int_{-1}^1 P_N(\theta)\,d\theta\int_\theta^1 P_n(\theta')\,
 w(\theta'-\theta)\,d\theta'.
\end{equation}
%
The notation
%
\begin{equation}
 q(\theta)=q_0w(\theta)
\end{equation}
%
is used here and later to separate the wake strength from its form, 
and to get the relation 
%
\begin{equation}
 R_{0,0}=\frac{1}{2}\int_0^2(2-\theta)\,w(\theta)\,d\theta = 1.
\end{equation}
%
Besides we will use the designations
%
\begin{subequations}
\begin{eqnarray}
 Z_n=\sum_m S_{n,m}C_{n,m},
\end{eqnarray}
\begin{eqnarray}
 W_n(\nu)=\sum_m\frac{|S_{n,m}|^2}{\nu-\nu_{n,m}}.
\end{eqnarray}
\end{subequations}
%
Then series~(14) obtains the most compact form
%
\begin{equation}
 Z_N=q_0 W_N\sum_{n=0}^\infty R_{N,n}Z_n. 
\end{equation}
%

%

\section{Constant wake}

%
%
\begin{table}[b!]
\begin{center}
\caption{Fragment of the matrix $R_{N,n}$. 
Its general form is given by Eq.~(13) and (16).}
\vspace{5mm}
\begin{tabular}{|c|c|c|c|c|c|c|}
\hline 
$n\rightarrow$ &    0    &    1    &    2     &   3      &    4   &   5    \\
\hline
$  ~ N=0 ~   $ &    1    &  ~1/3~  &     0    &   0      &    0   &~~~0~~~~\\
$  ~ N=1 ~   $ &~$-1/3$~ &    0    &    1/15  &   0      &    0   &~~~0~~~~\\
$  ~ N=2 ~   $ &    0    &$-1/15$~~&     0    &~~1/35~~  &    0   &~~~0~~~~\\
$  ~ N=3 ~   $ &    0    &    0    &~$-1/35$~~&   0      &~~1/63~~&~~~0~~~~\\
$  ~ N=4 ~   $ &    0    &    0    &     0    &~$-1/63$~~&    0   &~~1/99~~\\
$  ~ N=5 ~   $ &    0    &    0    &     0    &   0      &~$-1/99$~~& 0   \\
\hline
\end{tabular}
\end{center}
\end{table}
%
Several realistic examples of the wake will be considered in Sec.~V. 
However, the simplest model of constant wake $\,w=1,\;q=q_0\,$ is preliminary 
investigated in this section to discover the main features of the effect.
Though the wakes are negative in the most cases \cite{NG}, 
the positive wake is possible as well and et was observed in practice 
\cite{HER,GER}.
Therefore both signs of the parameter $\,q_0\,$ are analyzed in the section. 

It is easy to verify that, at $\,w=1\,$ and $\,N\ne 0$, the matrix 
$\,R_{N,n}$ is
%
\begin{equation}
 R_{N,n}=-\frac{\delta_{N-1,n}}{(2N-1)(2N+1)}+\frac{\delta_{N+1,n}}{(2N+1)(2N+3)}.
\end{equation}
%
(its small fragment is shown in Table~I).

One can check as well that $\,\nu_{0,0}=0,\;\,S_{0,0}=1$, 
that is $\,W_0=1/\nu$.
Using these features, one can represent the solvability condition of series (19), 
that is the dispersion equation for the bunch eigentunes, 
in terms of an infinite continued fraction
%
\begin{eqnarray}
 \nu-q_0+\frac{(q_0/3)^2 W_1}{1+\frac{(q_0/15)^2 W_1W_2}
 {1+\frac{(q_0/35)^2W_2W_3}{1+\dots\dots\dots}}}=0.
\end{eqnarray}
%
This expression has to be truncated in reality by applying of the assumption 
$\,W_n=0\,$ at $\,n\ge n_{\rm max}$. 
Assigning the truncated left-hand part of Eq.~(21) as $\,T_{n_{\rm max}}(\nu)$, 
one can write the approximate dispersion equation as
%
\begin{eqnarray}
 T_{n_{\rm max}}(\nu)=0
\end{eqnarray}
%
with the following recurrent relations:
%
\begin{eqnarray}
 T_n=T_{n-1}+T_{n-2}\frac{q_0^2\,W_{n-1}W_n}{(4n^2-1)^2}, \qquad (n\ge2)
\end{eqnarray}
%
and the boundary conditions:
%
\begin{subequations}
\begin{eqnarray}
 T_0=\nu-q_0,
\end{eqnarray}
\begin{eqnarray}
\qquad T_1 = \nu-q_0+\left(\frac{q_0}{3}\right)^2\frac{3(\nu+\Delta Q)}
 {\nu(\nu+\Delta Q)-Q_s^2}.
\end{eqnarray}
\end{subequations}

%

\subsection{Three-mode approximation}

%

Equation~(22) is trivial at $n_{\rm max}=0$: $\,T_0)\nu)=0\,$ that is $\,\nu=q_0$,
as it follows from Eq.~(24a), 
It describes the wake contribution to the tune of the lowest (rigid) 
head-tail mode of the bunch.
Of course, the TMCI cannot appear in this approximation, and the simplest equation  
to disclose it is $\,T_1(\nu)=0$ that is, in accordance with Eq.~(24b)
%
\begin{eqnarray}
 (\nu-q)\left(\nu-\frac{Q_s^2}{\nu+\Delta Q}\right)=-\frac{q_0^2}{3}.
\end{eqnarray}
%
This third order equation exactly coincides with Eq.~(7.3) of 
Ref.~\cite{BA2} (without chromaticity) despite the fact that the very 
different concepts have been used to derive them.
However, the examples presented in \cite{BA2} have been restricted by 
the modest SC tune shift: $\Delta Q/Q_s<3$. 
The situation beyond this region will be explored here for the best understanding 
of the phenomenon, 
and for further development of the techniques.
%
 \begin{figure}[t!]
 \includegraphics[width=85mm]{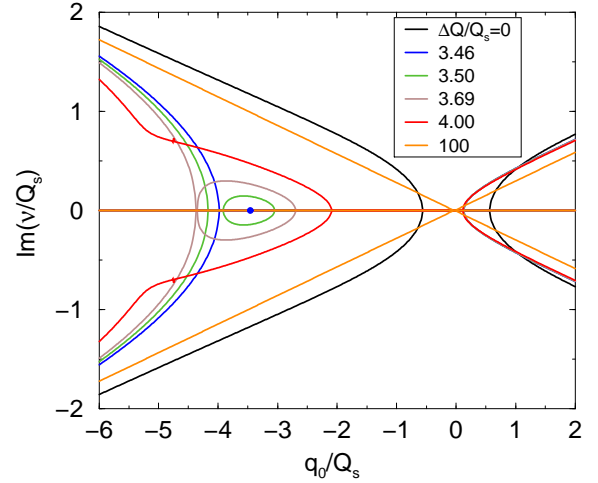}
 \caption{Imaginary part of the boxcar eigentunes
 against the wake strength at different value of space charge tune shift. 
 There are 2 regions of instability if the wake is negative 
 and $\,3.46<\Delta Q/Q_s<3.69\,$.}
 \end{figure}
%
 \begin{figure}[h!]
 \includegraphics[width=85mm]{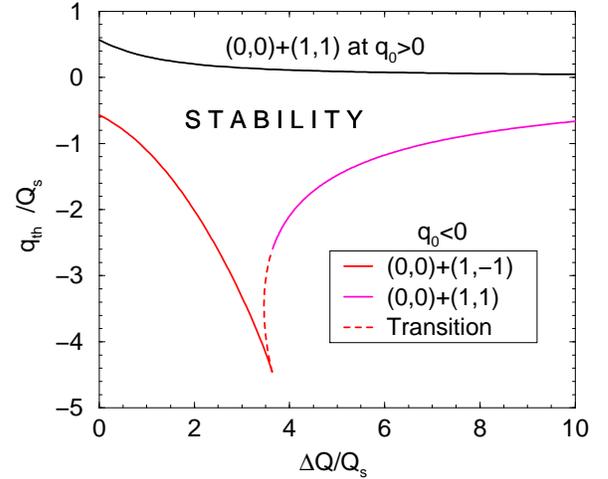}
 \caption{Instability threshold of the boxcar bunch against SC tune shift
 (positive and negative wakes). Indexes of the coalesced modes are shown 
 for each part of the threshold line. }
 \end{figure}
%

Imaginary part of a solution of Eq.~(25) is plotted in Fig.~3 against the wake 
strength at different SC.
According to the plot, the instability threshold is $\,|q_{\rm th}/Q_s|=0.567\,$ 
at $\,\Delta Q=0\,$ (the black lines). 
Its dependence on $\,\Delta Q/Q_s\,$ is shown in Fig.~4.
The plot is very simple at $\,q_0>0$: 
the threshold goes down monotonically tending to 0 when the space charge increases. 
The case of negative wake is more complicated and requires a special comment.
Absolute value of its threshold increases with $\,\Delta Q\,$ reach[nd
$\,q_{\rm th}/Q_s\simeq -4\,$ at $\,\Delta Q/Q_s=3.46$ (blue parabola in Fig.~3).
The picture crucially changes after that because a new region of instability arises 
whose initial position is shown as the blue point in Fig.~3. 
Then it quickly expands stepping through the green oval to the brown one
and joining with the primary region of instability (brown parabola) 
at $\,\Delta Q/Q_s=3.69$.
The barrier between the parts tears at higher $\,\Delta Q\,$ resulting in a single 
region of instability (red).
Its right-hand boundary goes to the right, that is the TMCI threshold goes down,
if the SC tune shift continues to grow up.

%

\subsection{Higher approximations}

%
Higher approximations should be involved to validate the three-mode model,
to establish its applicability limit, and to go beyond it.

The first step in this way is an investigation of the equation $\,T_2(\nu)=0$.
According to Eqs.~(23) and (24), its expanded form  is
%
\begin{eqnarray}
 \nu-q_0+\frac{q_0^2\,W_1(\nu)}{9}
+\frac{q_0^2\,(\nu-q_0)W_1(\nu)W_2(\nu)}{225}  = 0
\end{eqnarray}
%
where
%
\begin{subequations}
\begin{equation}
 W_1(\nu)=\frac{3(\nu+\Delta Q)}{\nu(\nu+\Delta Q)-Q_s^2},\qquad 
\end{equation}
\begin{equation}
 W_2(\nu)=\frac{|S_{2,-2}|^2}{\nu-\nu_{2,-2}}+\frac{|S_{2,0}|^2}{\nu-\nu_{2,0}}
+\frac{|S_{2,2}|^2}{\nu-\nu_{2,2}}.
\end{equation}
\end{subequations}
%
Required parameters have to be obtained by solution of Eq.~(13) with $\,n=2$ 
as it is described in the Appendix.
With the notations $\,\nu_{n,m}=\hat\nu_{n,m}Q_s-\Delta Q$,
the eigentunes appear as all roots of the dispersion equation
%
\begin{equation}
 \hat\nu_{2,m}(\hat\nu_{2,m}^2-4)=\frac{\Delta Q}{Q_s}\,(\hat\nu_{2,m}^2-1),
\end{equation}
%
and formula for corresponding normalizing coefficients is
%
\begin{equation}
 |S_{2,m}|^2=\frac{5(\hat\nu_{2,m}^2-1)^2}{\hat\nu_{2,m}^4+\hat\nu_{2,m}^2+4}.
\end{equation}
%
The substitution of the functions $\,W_{1-2}(\theta)\,$ into Eq.~(26) results in 
the equation of $6^{\rm th}$ power
%
\begin{eqnarray}
 (\nu-q_0)\left(\nu-\frac{Q_s^2}{\nu+\Delta Q}\right)+\frac{q_0^2}{3}
=-\frac{q_0^2(\nu-q)}{75} \nonumber \\
 \times\left(\frac{|S_{2,-2}|^2}{\nu-\nu_{2,-2}}
 +\frac{|S_{2,0}|^2}{\nu-\nu_{2,0}}+\frac{|S_{2,+2}|^2}{\nu-\nu_{2,+2}}\right)
 \!\!.
\end{eqnarray}
%
This equation has 6 roots which are different real numbers inside the 
stability area.
However, at least 2 of them should be coinciding at the boundary of this area,
which feature can be used for the search of the instability threshold.
Corresponding threshold of negative wake is presented in Fig.~5 
by the magenta line. 
The case $\,q_0>0\,$ is not considered in this subsection 
because the result almost does not depend on $\,n_{max}\,$
and can be reasonably described by the three-mode approximation, Eq.~(25).
%
\begin{figure}[t!]
 \includegraphics[width=85mm]{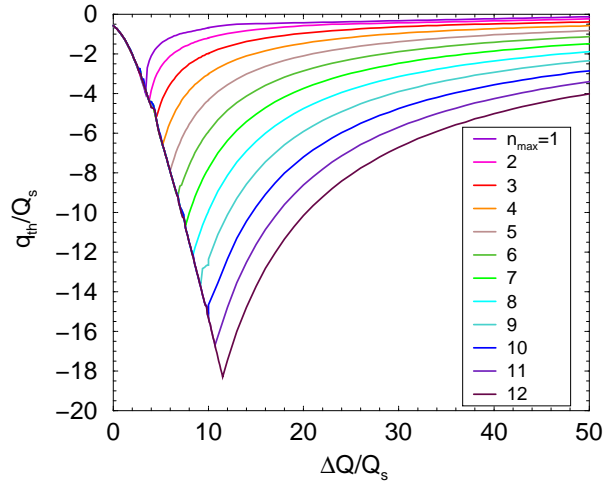}
 \caption{Threshold curve of the boxcar bunch in different approximations.
 Index $\,n_{\rm max}\,$ means maximal power of the Legendre polynomial 
 in the expansion.} 
\end{figure}
%

Similar method can be used for analysis of higher approximations 
though corresponding formulae are essentially more cumbersome.
Generally, the case involves $\,(n_{\rm max}+1)(n_{\rm max}+2)/2\;$ 
basis vectors and leads to algebraic equation of the same power, 
where $\,n_{\rm max}\,$ is order of the highest used Legendre polynomial.

Results of the calculations are collected in Fig.~5 at $\,n_{max}=1-12$
(dispersion equation of $3-91$ power).
It is seen that, at rather small SC, absolute value of the threshold 
rises with $\,\Delta Q$, different approximations provide actually 
coinciding results in their region of applicability, and each additional step 
simply expands this region.
For example, the three-mode approximation ($n_{max}=1$) provides  correct 
magnitude of the threshold at $\,\Delta Q/Q_s\le 3.46\,$ but at least 
$\,n_{max}=12\,$ (91-mode approximation) is needed to get proper results within 
the range $\,\Delta Q/Q_s=0-12$.

The sequential decrease of the threshold cannot be treated as a physical effect
because of absence of the convergence.
The opposite assumption which has been admitted in my preprint \cite{BA3} 
was coming from an insufficient accuracy of numerical calculations
which has led to an incomplete separation of numerous and very tightly spaced 
radial modes.

%

\subsection{The bunch spectrum}

%

The inadequate convergence of the curves in Fig.~5 remains the open question:
which is the TMCI threshold of negative wake at very large magnitude 
of $\,\Delta Q/Q_s$, such as several tens or more?
There is a related information in Ref.~\cite{BA1}:
at such conditions, the TMCI cannot be caused by a coalescence of positive 
eigentunes of the bunch.
The last reservation is important because only a part of the boxcar modes 
has been used for the analysis in \cite{BA1}. 
Tunes of these modes are located in the the upper part of Fig.~1.
Therefore, more detailed examination of the bunch spectrum is needed with the wake
to check the results, including all bunch tunes $\,\nu_{th}(\Delta Q)$ at the 
frontier of the TMCI area.

%
 \begin{figure}[t!]
 \includegraphics[width=85mm]{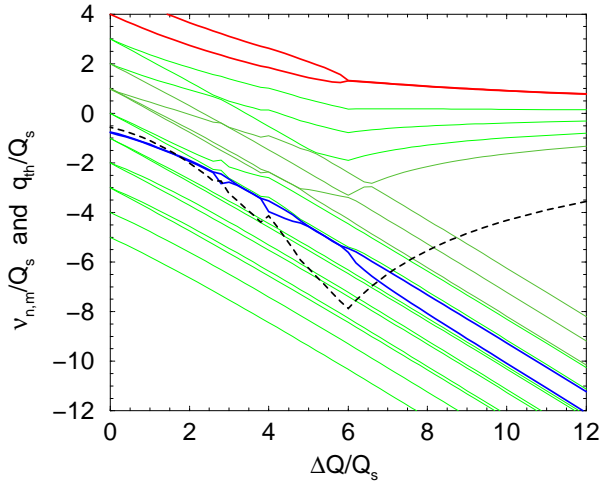}
 \caption{The bunch spectrum in the TMCI frontier at $\,n_{\rm max}=5\,$ 
 (21-mode approximation). 
 The most important modes are shown by blue and red lines.
 The TMCI threshold is represented by the dashed black line.} 
 \end{figure}
%

Very first example of this has been given in Fig.~4 where indexes of the 
coalesced modes are specified for the three-mode approximation.
The more detailed view is represented in Fig.~6 where full spectrum of the bunch
is plotted at $\,n_{max}=5\,$ (21-mode approximation).
The most important spectral lines are displayed by special colors: 
blue for the modes $\{0,0\}\,$ and $\,\{1,-1\}$,
and red for the modes $\{5,5\}$ and $\{4,4\}$. 
According to the picture, the coalescence of these modes is responsible for the 
TMCI at $\,\Delta Q/Q_s<6\,$ or $>6$, correspondingly. 
Just the switching from the lower pair to the upper one causes the sharp kink 
of the threshold curve which is shown in the plot by dashed line.
An incidental interference of other modes slightly affects the curve
but does not change its general contour.
%
 \begin{figure}[t!]
 \includegraphics[width=85mm]{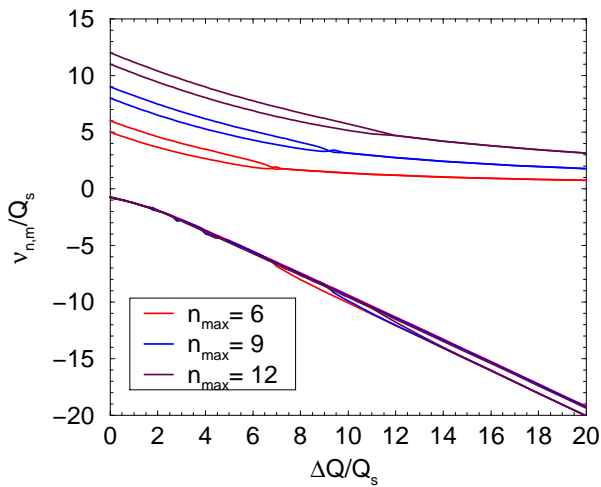}
 \caption{The most important spectral lines in different approximations.
 The lower lines: modes $\{0,0\}\,$ and $\,\{1,-1\}$; the upper ones: 
 the highest observed tunes in given approximation.}
 \end{figure}
%

Another examples are given in Fig.~7 where the most important spectral lines 
are plotted at three different approximations with  $\,n_{max}=6,$~9,~12.
The lower curves represent tunes of the modes $\,\{0,0\}+\{1,-1\}\,$
which are coalesced in the beginning and are closely located later.
The convergence manifests itself in the fact that the coalesced part
of the curves expands when $\,n_{max}\,$ increases. 

The upper curves in Fig.~7 represent the tunes 
$$
  \{n_{max},n_{max}\}+\{n_{max}-1,n_{max}-1\} 
$$
It is seen that the lines of the same color merge at rather large 
$\,\Delta Q/Q_s\,$.
However, it is a divergent process because the junction point does not tend 
to a certain limit when $\,n_{max}$ increases.
It inevitably leads to the conclusion that the junction of the positive tunes
in Figs.~6 and 7, as well as the associated leap of the threshold, are not a 
physical effects.
The engaging of Ref.~\cite{BA1} allow to assert that this statement should 
be true in any approximation

Therefore a monotonous rising of the TMCI threshold looks as the most 
credible assumption.
It also compliance with behavior of the low crucial modes which are 
$$
\nu_{0,0}\sim q_0,\quad\nu_{1,-1}\simeq -Q_s-\Delta Q.
$$
According this,  the instability condition $\,\nu_{0,0}=\nu_{1,-1}\,$ 
should result in the threshold relation $\,q_{th}\sim-\Delta Q$.

%

\section{TMCI with realistic wake}

%

Series of equations (19) with matrix $\,R_{N,n}\,$ given by Eq.~(15) 
is applicable at any wake function $\,q(\theta)=q_0 w(\theta)$. 
However, dispersion equation (21) and its approximate forms provided by 
Eqs.~(22)-(24) are valid only with constant wake.
Therefore, more standard procedures  should be generally used for solution 
of Eq.~(19). 
It results in some deterioration of the accuracy and compels to restrict
number of used basis vectors.
Performed calculations with constant wake attest that the ``old'' and the ``new'' 
results coincide at $\,n_{max}\le 9\,$ (55-modes approximation),
otherwise some real roots can be lost.  
This restriction is accepted below at the calculations with realistic wakes.

%

\subsection{Resistive wall wake}

%

Resistive wall impedance is the most general and important source of transverse 
instabilities in circular accelerators. 
At $\,z\gg\b/\gamma\,$,Its wake function is \cite{NG}:

%
\begin{table}[b!]
\begin{center}
\caption{Fragment of the matrix $R_{N,n}$ for resistive wake.}
\vspace{5mm}
\begin{tabular}{|c|c|c|c|c|c|c|}
\hline 
$n\rightarrow$ &    0   &    1   &    2   &   3    &    4   &   5    \\
\hline
$  ~ N=0 ~   $ &    1   & .20000 &-.02857 & .00952 &-.00433 & .00233 \\
$  ~ N=1 ~   $ &-.20000 & .14286 & .06667 &-.01299 & .00513 &-.00260 \\
$  ~ N=2 ~   $ &-.02857 &-.06667 & .06494 & .03590 & .00779 & .00332 \\
$  ~ N=3 ~   $ &-.00952 &-.01299 &-.03590 & .03896 & .02323 &-.00533 \\
$  ~ N=4 ~   $ &-.00433 &-.00513 &-.00779 &-.02323 & .02666 & .01659 \\
$  ~ N=5 ~   $ &-.00233 &-.00260 &-.00332 &-.00533 &-.01659 & .01970 \\
\hline
\end{tabular}
\end{center}
\end{table}
%
%
 \begin{figure}[t!]
 \includegraphics[width=85mm]{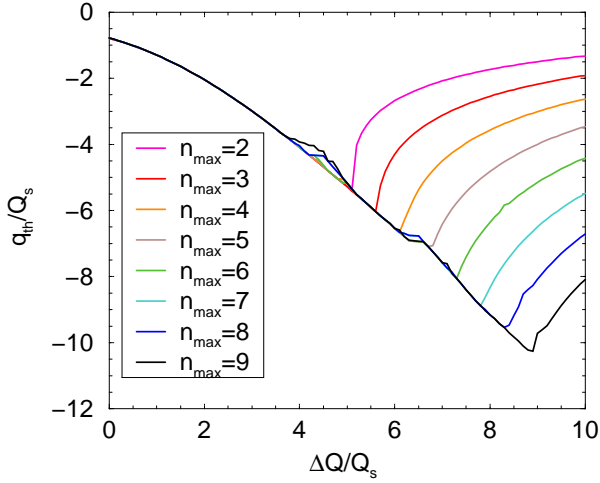}
 \caption{Stability region of the boxcar bunch with resistive wall wake.
 Normalized magnitude of the wake is given by Eq.~(32),
 index $\,n_{\rm max}\,$ means maximal power of Legendre polynomial in the 
 expansion.} 
\end{figure}
%
 \begin{equation}
 W_1(z) = -\frac{4R}{b^3}\sqrt{\frac{c}{\sigma z}}
 \end{equation}
%
where $\,b\,$ is the beam pipe radius, $\,\sigma\,$ is the pipe wall conductivity,
and $\,z\,$ is distance from the field source to the observation point. 
Taking into account Eqs.~5 and (17), one can write the normalized wake function as
$\,q=q_0w(\theta)$ with
%
\begin{eqnarray}
 q_0 = -\frac{4 r_0R^2N_b}{3\pi\gamma\beta b^3Q_0}\sqrt{\frac{c}{\sigma z_b}},
 \qquad w(\theta) = \frac{3}{4\sqrt{2\theta}}
\end{eqnarray}
%
where $\,z_b\,$ is the bunch length in usual units.
Generally speaking, this wake can cause multiturn collective effects as well.
However, their influence on the TMCI threshold is negligible for a single bunch 
with $\,z_b\ll 2\pi R\,$ \cite{BA2} which assumption is used further.
Then series (19) is applicable with the matrix $\,R_{N,n}\,$
whose part is represented in Table II.
Threshold of this instability is plotted in different approximations in Fig.~8.
which is very similar to Fig.~5 (constant wake) both in the form and in the magnitude.

%

\subsection{Short square wake}

%

A square wake can be created by a strip-line BPM or by a traveling-wave kicker 
\cite{NG}.
The long (constant) square wake is considered above in detail.
However, the wake can be shorter than the bunch, in practice.
In accordance with Eq.~(17), its normalized strength should be represented 
in the form 
%
\begin{equation}
 q(\theta) = q_0w(\theta)=\frac{4q_0}{\theta_w(4-\theta_w)}
 \quad {\rm at}\quad 0<\theta<\theta_w
\end{equation}
%
where $\,\theta_w<2\,$ is the wake length 
(recall that the bunch length is 2 in these units).  
Several examples are represented in Fig.~9 at $\,n_{max}=9$. 
Note that the horizontal lines have no physical sense and are added
to mark end of the curve applicability 
(the calculations were not carried out after that).
Due to the normalization, the threshold dependence on the SC tune shift is 
not very significant, especially at $\Delta Q/Q_s<5$.
%
 \begin{figure}[h!]
 \includegraphics[width=85mm]{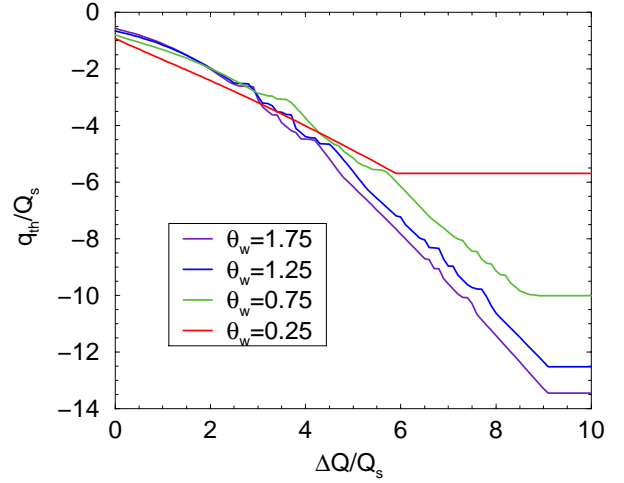}
 \caption{TMCI threshold of short square wake against SC tune shift. 
 Length of the wake is $\,\theta_w$, the bunch length=2.
 The approximation with $\,n_{\max}=9\,$ (55 modes) is used, the horizontal 
 lines mark end of the applicability area.}
 \end{figure}
%

%

\subsection{Oscillating wake}

%
%
 \begin{figure}[b!]
 \includegraphics[width=85mm]{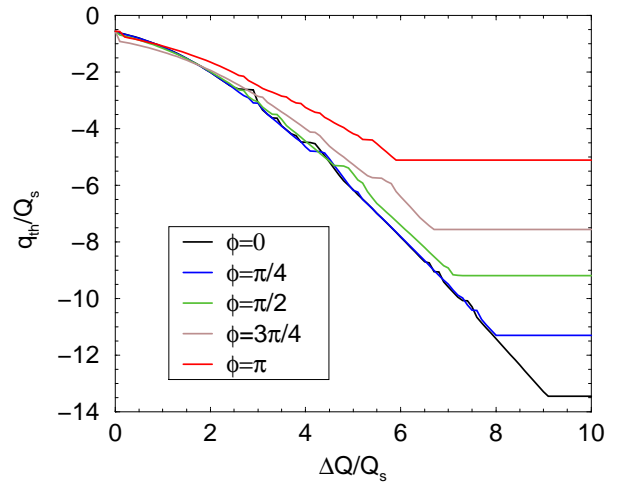}
 \caption{TMCI threshold of oscillating wake with phase advance $\,\phi\,$
 within the bunch of length $\,\Delta\theta = 2$. $n_{max}=7\,$ 
 (36-mode approximation).}
 \end{figure}
%
There are several models considering the wake field source as a resonator 
of frequency $\,f=c/\lambda$ \cite{NG}. 
It creates an oscillating wake $\propto\cos(2\pi z/\lambda)$ 
having the phase advance $\,\phi=2\pi z_b/\lambda\,$ within the bunch.
We consider the case $\,\phi<2\pi\,$, and represent  
the wake in the form satisfying the normalization condition (17):
%
\begin{equation}
 q(\theta) = q_0w(\theta),\qquad w(\theta)=\frac{\phi^2\cos(\phi\theta/2)}
 {2(1-\cos\phi)}
\end{equation}
%
Several examples are represented in Fig.~10 at $\,n_{max}=7$
to demonstrate that SC produces a stabilizing effect in these cases as well.

%

\section{Conclusion}

%

Being stable in themselves, the eigenmodes of the boxcar bunch with space charge
form a convenient and effective basis for investigation of the bunch instability 
with the space charge and a wake field.	 
The dispersion equation derived by this method at the constant wake
is represented in the form of an infinite continued fraction as well as in the form 
of a recursive relation with arbitrary number of the basis functions involved.

It is shown that the TMCI threshold of the negative constant wake grows up 
in absolute value when the SC tune shift increases.
An enlargement of number of the used basis vectors expands area of 
applicability of this statement but does not change the results obtained before.
The statement is confirmed in the paper by the straightforward calculation 
of the threshold at $\,\Delta Q/Q_s\le 12\,$ using the basis set including 
up to 91 eigenfunctions.
Convergence of the solutions is not achieved at higher SC 
because very large number of the eigenfunctions is required for 
separation of different radial modes.
However, an additional analysis of the bunch spectrum allows to extend the 
statement to any tune shift.

Similar results are obtained with realistic wake functions including the resistive 
wall, the short square, and the oscillating forms.

Threshold of the positive wake goes down when the SC tune shift increases,
and the effect can be satisfactory described by the three-mode approximation.


\section{Acknowledgments}

FNAL is operated by Fermi Research Alliance, LLC under contract 
No. DE-AC02-07CH11395 with the United States Department of Energy.


\section{Appendix}

%
Using the notation
$$
 \hat\nu_{n,m}=\frac{\nu_{n,m}+\Delta Q}{Q_s},\quad
 \hat\Delta Q =\frac{\Delta Q}{Q_s},\quad
 P_n(\theta) = \sum_{l=0}^n p_{nl}\theta^l
 \eqno(A1)
$$
one can rewrite Eq.~(13) in the form
$$
 \hat\nu_{n,m}Y_{n,m}+i\,\frac{\partial Y_{n,m}}{\partial\phi}=
 S_{n,m}\Delta\hat Q \sum_{l=0}^np_{n,l}(A\cos\phi)^l
 \eqno(A2)
$$
Its solution is
$$
 Y_{n,m}=S_{n,m}\Delta\hat Q\sum_{k=-n}^n\frac{\exp{ik\phi}}{\hat\nu_{n,m}-k}
 \sum_{j=0}^n U_{n,k,j}A^{k+2j} 
 \eqno(A3)
$$
where
$$
 U_{n,k,j}=\frac{p_{n,k+2j}}{2^{k+2j}}{k+2j\choose j} \times
 \bigg\{\begin{array}{ll} 1 {\rm\;\;at}\;\;k+j\ge 0     \\
                          0 {\rm\;\;at}\;\;k+j <  0     \end{array}
 \eqno(A4)
$$
This function should satisfy normalization condition represented by Eq.~(7) with 
$\,j\equiv\{n,m\}\,$ and distribution function (11a).
The substitution results in the relation: 
$$
 \frac{1}{S_{n,m}^2\Delta\hat Q^2} =\sum_{k=-n}^n\frac{1}{(\hat\nu_{n,m}-k)^2}
$$
\vspace{-3mm}
$$
 \times\sum_{j_1=0}^n\sum_{j_2=0}^n U_{n,k,j_1} U_{n,k,j_2}
 \overline{A^{2(k+j_1+j_2)}}
 \eqno(A5)
$$
where $\,\overline {A^{2j}}\,$ is the amplitude power averaged over the 
distribution:
$$
 \overline{A^{2j}}=\int_0^1\frac{A^{2j+1}\,dA}{\sqrt{1-A^2}}=
 \sum_{l=0}^j {j\choose l} \frac{(-1)^l}{2l+1}
 \eqno(A6)
$$
In principle, involved eigentunes $\,\nu_{n,m}\,$ could be obtained by 
substitution of Eq.~(A3) into Eq.~(4) with the functions $\,Y\,$ and $\,\bar Y\,$
being taken from this Appendix.
Because similar calculation has been actually accomplished in Ref.~\cite{SAH}, we
represent here only the resulting equation for the eigentunes:
\\
\\
Lower powers
$$
 \hat\nu_{0,0}=\Delta\hat Q,\qquad 
 \hat\nu_{1,\pm1}^2-1=\Delta\hat Q\hat\nu_{1,\pm1}
 \eqno(A7)
$$
Higher even powers
$$
 \hat\nu_{n,m}[\hat\nu_{n,m}^2-4]\dots[\hat\nu_{n,m}^2-n^2]
$$
$$
=\Delta\hat Q [\hat\nu_{n,m}^2-1]\dots[\hat\nu_{n,m}^2-(n-1)^2]
\eqno(A8)
$$
Higher odd powers
$$
 [\hat\nu_{n,m}^2-1]\dots[\hat\nu_{n,m}^2-n^2]
$$
$$
=\Delta\hat Q\hat\nu_{n,m} [\hat\nu_{n,m}^2-4]\dots[\hat\nu_{n,m}^2-(n-1)^2]
\eqno(A9)
$$
\\
Some of the values $\,\nu_{n,m}/Q_s=\hat\nu_{n,m}-\Delta\hat Q\,$ are plotted 
in Fig.~1.
Factors $S_{n,m}^2$ can be found from Eq.~(A5) with the known eigentunes 
substituted. 
Some results are represented below:
$$
 S_{0,0}^2=1,\qquad S_{1,\pm1}^2=\frac{3\hat\nu_{1,\pm1}^2}{\hat\nu_{1,\pm1}^2+1},
\eqno(A10)
$$ 
$$
 S_{2,m}^2=\frac{5(\hat\nu_{2,m}-1)^2}
 {\hat\nu_{2,m}^4+\hat\nu_{2,m}^2+4},\quad m=2,\,0,-2,
\eqno(A11) 
$$
$$
 S_{3,m}^2=\frac{7\hat\nu_{3,m}^2(\hat\nu_{3,m}^2-4)^2}
 {\hat\nu_{3,m}^6-2\hat\nu_{3,m}^4+13\hat\nu_{3,m}^2+36},
 \eqno(A12) 
$$
etc.
%


\end{document}